\documentclass[aps,superscriptaddress,amsmath,amssymb,floatfix,twocolumn,showpacs,amsfonts,longbibliography,prl]{revtex4-1}
\usepackage{times}
\usepackage[varg]{txfonts}
\usepackage{textcomp}
\usepackage{graphicx}
\usepackage{subfigure}
\usepackage{tabu}
\usepackage{color}
\usepackage[colorlinks=true,citecolor=blue,urlcolor=blue,linkcolor=blue,hyperindex]{hyperref}
\usepackage{braket}
\usepackage{float}
\usepackage{overpic}
\usepackage{gensymb}
\usepackage{mathrsfs}
\usepackage{tikz}
\usepackage{bm}
\usepackage{multirow}
\usepackage{amsfonts}
\usepackage{amsmath}
\usepackage{mathrsfs}
\usepackage{amssymb}
\usepackage{paralist}
\usepackage{indentfirst}
\usepackage{ulem}
\usepackage{soul}
\usepackage{cancel}
\usepackage{CJK}

\allowdisplaybreaks

\begin{document}
\title{Non-integer Floquet Sidebands Spectroscopy}
\author{Du-Yi Ou-Yang}
\affiliation{Department of Physics, and Chongqing Key Laboratory for Strongly Coupled Physics, Chongqing University, Chongqing, 401331, China}
\affiliation{Center of Modern Physics, Institute for Smart City of Chongqing University in Liyang, Liyang, 213300, China}

\author{Yan-Hua Zhou}
\affiliation{Department of Physics, and Chongqing Key Laboratory for Strongly Coupled Physics, Chongqing University, Chongqing, 401331, China}
\affiliation{Center of Modern Physics, Institute for Smart City of Chongqing University in Liyang, Liyang, 213300, China}

\author{Ya Zhang}
\affiliation{School of Instrumentation Science and Engineering, Harbin Institute of Technology, Harbin, 150000, China}

\author{Xiao-Tong Lu}
\affiliation{Key Laboratory of Time and Frequency Primary Standards, National Time Service Center, Chinese Academy of Sciences, Xi'an 710600, China}
\affiliation{School of Astronomy and Space Science, University of Chinese Academy of Sciences, Beijing 100049, China}

\author{Hong Chang}
\affiliation{Key Laboratory of Time and Frequency Primary Standards, National Time Service Center, Chinese Academy of Sciences, Xi'an 710600, China}
\affiliation{School of Astronomy and Space Science, University of Chinese Academy of Sciences, Beijing 100049, China}

\author{Tao Wang}
\thanks{corresponding author:  tauwaang@cqu.edu.cn}
\affiliation{Department of Physics, and Chongqing Key Laboratory for Strongly Coupled Physics, Chongqing University, Chongqing, 401331, China}
\affiliation{Center of Modern Physics, Institute for Smart City of Chongqing University in Liyang, Liyang, 213300, China}

\author{Xue-Feng Zhang}
\thanks{corresponding author:  zhangxf@cqu.edu.cn}
\affiliation{Department of Physics, and Chongqing Key Laboratory for Strongly Coupled Physics, Chongqing University, Chongqing, 401331, China}
\affiliation{Center of Quantum Materials and Devices, Chongqing University, Chongqing 401331, China}

\begin{abstract}
In the quantum system under periodical modulation, the particle can be excited by absorbing the laser photon with the assistance of integer Floquet photons, so that the Floquet sidebands appear. Here, we experimentally observe non-integer Floquet sidebands (NIFBs) emerging between the integer ones while increasing the strength of the probe laser in the optical lattice clock system. Then, we propose the Floquet channel interference hypothesis (FCIH) which surprisingly matches quantitatively well with both experimental and numerical results. With its help, we found both Rabi and Ramsey spectra are very sensitive to the initial phase and exhibit additional two symmetries. More importantly, the height of Ramsey NIFBs is comparable to the integer one at larger $g/\omega_s$ which indicates an exotic phenomenon beyond the perturbative description. Our work provides new insight into the spectroscopy of the Floquet system and has potential application in quantum technology.
\end{abstract}
\maketitle

\textbf{\textit{Introduction.--}} Floquet theory provides an ideal bridge between the equilibrium and non-equilibrium quantum system. Under periodical modulation with driving frequency $\omega_s=2\pi/T_s$, the time evolution can be well described by a time-independent effective Hamiltonian in certain conditions \cite{Eckardt_2007}. Consequently, the scope of controlling and manipulating the quantum system is widely extended, and this so-called `Floquet engineering' becomes a conventional tool for tailoring exotic Hamiltonian \cite{EckardtRMP2017}. The Floquet theory can be taken as an analog of the Bloch theory in the time dimension, and the $k$-th order effective Floquet Hamiltonian corresponds to the Floquet channel at $k\omega_s$ in the frequency space. 

\begin{figure}[t]
	\centering
	\includegraphics[width=0.99\linewidth]{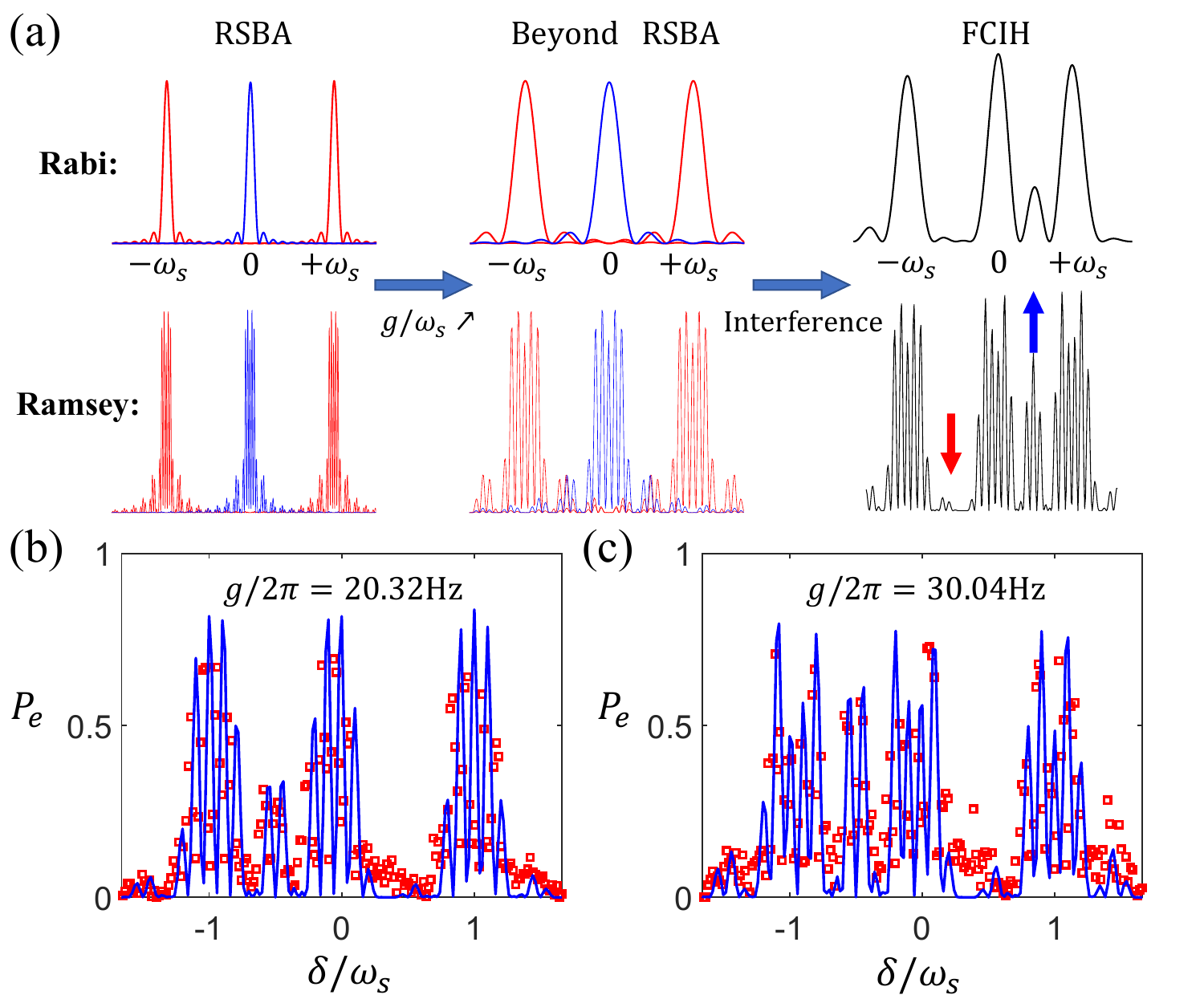}
	\caption{(a) The schematic diagrams of Rabi and Ramsey spectrum explaining the mechanism of NIFBs caused by interference of different Floquet channels. (b) and (c) shows the experimental (red dots) and numerical (blue lines) results of the Ramsey spectrum with different $g$ at $\phi=\pi$ which exchanges the positive and negative NIFBs.
 }\label{fig1}
\end{figure}

The typical example is the Rabi model which describes the interplay between light and matter, and is strongly related to various important platforms of quantum sensing and quantum control \cite{geometric_zeng2019,observation_cai2021}, such as NV center \cite{NV_center_Dobrovitski2013,NV_center_golter2016}, trapped ion \cite{trapped_ions_wineland2003,trapped_ions_blatt2012}, optical lattice clock \cite{Blatt}, etc. If the frequency of the light $\omega_p(t)$ is periodically modulated, the Hamiltonian within the rotating-wave-approximation (RWA) can be written as
\begin{eqnarray}\label{eq1}
\hat{H}=\frac{\hbar}{2}[(\omega_0-\omega_p(t))\hat{\sigma}_z+g\hat{\sigma}_x].
\end{eqnarray} 
where $\hat{\sigma}_{z(x)}$ is the Pauli matrix labeling the longitudinal (transversal) interaction of the two-level atom with transition frequency $\omega_0$ and $g$ is the Rabi frequency. The driving form is $\omega_p(t)=\bar{\omega}_p-A\omega_s\cos\left(\omega_s t+\phi\right)$ in which $\bar{\omega}_p$ is the mean-value, $A$ is the driving amplitude, and $\phi$ is the initial phase. 
Then, with the definition of the detuning $\delta=\omega_0-\bar{\omega}_p$, we can find the carrier peak is split to several Floquet sidebands at $\delta=k\omega_s,k\in\mathbb{Z}$ as shown in Fig.\ref{fig1}(a). When $g \ll \omega_s$, with the help of high-frequency expansion, each Floquet sideband can be described by a different time-independent Rabi model, which is known as resolved sideband approximation (RSBA) \cite{Blatt,MoJuanYin2020,XiaoTongLu2020}. However, when $g/\omega_s$ increases, both nearest neighboring Floquet sidebands will exert a strong influence over the physics in the intermediate region. Then, the \textit{non-integer Floquet sidebands} (NIFBs) could emerge in both Rabi and Ramsey spectroscopy as depicted in Fig.\ref{fig1} (a).

In this manuscript, as shown in Fig.\ref{fig1} (b-c), the NIFBs are experimentally observed in the optical lattice clock (OLC) platform. Both numerical and experimental results demonstrate they are very sensitive to the initial phase of the periodic driving. To understand these exotic phenomena, we propose the \textit{Floquet channel interference hypothesis} (FCIH) which works well at small $g/\omega_s$. The NIFBs can be strongly enhanced by the interference, and their height can be even comparable to the integer ones in the Ramsey spectroscopy (see Fig.\ref{fig1}(c)). At last, the relation between the initial phase and NIFBs is experimentally verified.

\textbf{\textit{Methods--}} In the experiment, after a two-stage Doppler cooling process, about $10^4$ $^{87}$Sr atoms are cooled down to $\approx 3 \mu$K and trapped in one-dimensional optical lattice \cite{Blatt,MoJuanYin2020,wang2018strontium}. The lattice laser is set to ``magic wavelength" so that the AC Stack shifts of both ground state ($^1$S$_0$) and excited state ($^3$P$_0$) are the same. The long lifetime of the excited state can make sure the spontaneous emission can be ignored \cite{XiaoTongLu2020}. The depth of the lattice potential is $V_0/Er\approx93$ ($Er=3.44$kHz is the recoil energy) so that the tunneling between nearest neighbor sites is strongly suppressed and each site can be taken to be irrelevant to the others. All the atoms are initially prepared in the ground state. Then, the clock laser is added to excite the atom from the ground state to the excited state. At the same time, its frequency is periodically modulated as $\omega_p(t)$ with $\omega_s/2\pi=100$Hz. The misaligned angle between the lattice and clock laser is $\delta\theta=0.008$. Meanwhile, the longitudinal and transverse trap frequencies are $\nu_z=64$kHz and $\nu_r=250$Hz, respectively \cite{MoJuanYin2020,zhou}. 
For the Rabi spectroscopy, the probing time is set to $3T_s$. On the other hand, in Ramsey spectroscopy, the dark time is $t_d=6T_s$ and the probing time is $t_p=3T_s$. Notice that, in order to avoid introducing additional discontinuity of the periodical driving, the clock laser is blocked out during the dark time. Following our previous work \cite{MoJuanYin2020,Xiaotong2022,XiaoTongLu2020,zhou}, the numerical results can be calculated with the help of the Runge-Kutta method and they quantitatively match very well with the experimental results in the deep optical lattice potential.


According to the FSBA, the $k-$th Floquet sideband can be understood as follows: the atom can be excited by absorbing the photon of the clock laser with the assistance of the $k-$th Floquet photons. As a consequence, the Rabi frequency is renormalized as $g_k=gJ_{-k}[A]e^{-ik\phi}$ where $J_k[]$ is the $k$th order first kind Bessel function, and the wave function contains additional phase $e^{\frac{-ik\omega_s t}{2}}$ which doesn't affect the excited population. However, the emergence of the NIFBs hints strong interference effect beyond FSBA, so we propose the FCIH as follows:
\begin{quote}
\textit{	The quantum dynamics of the two-level atom at small $g/\omega_s$ is approximately determined by the interference of all Floquet channels.}
\end{quote}

\begin{figure}[t]
	\centering
	\includegraphics[width=0.99\linewidth]{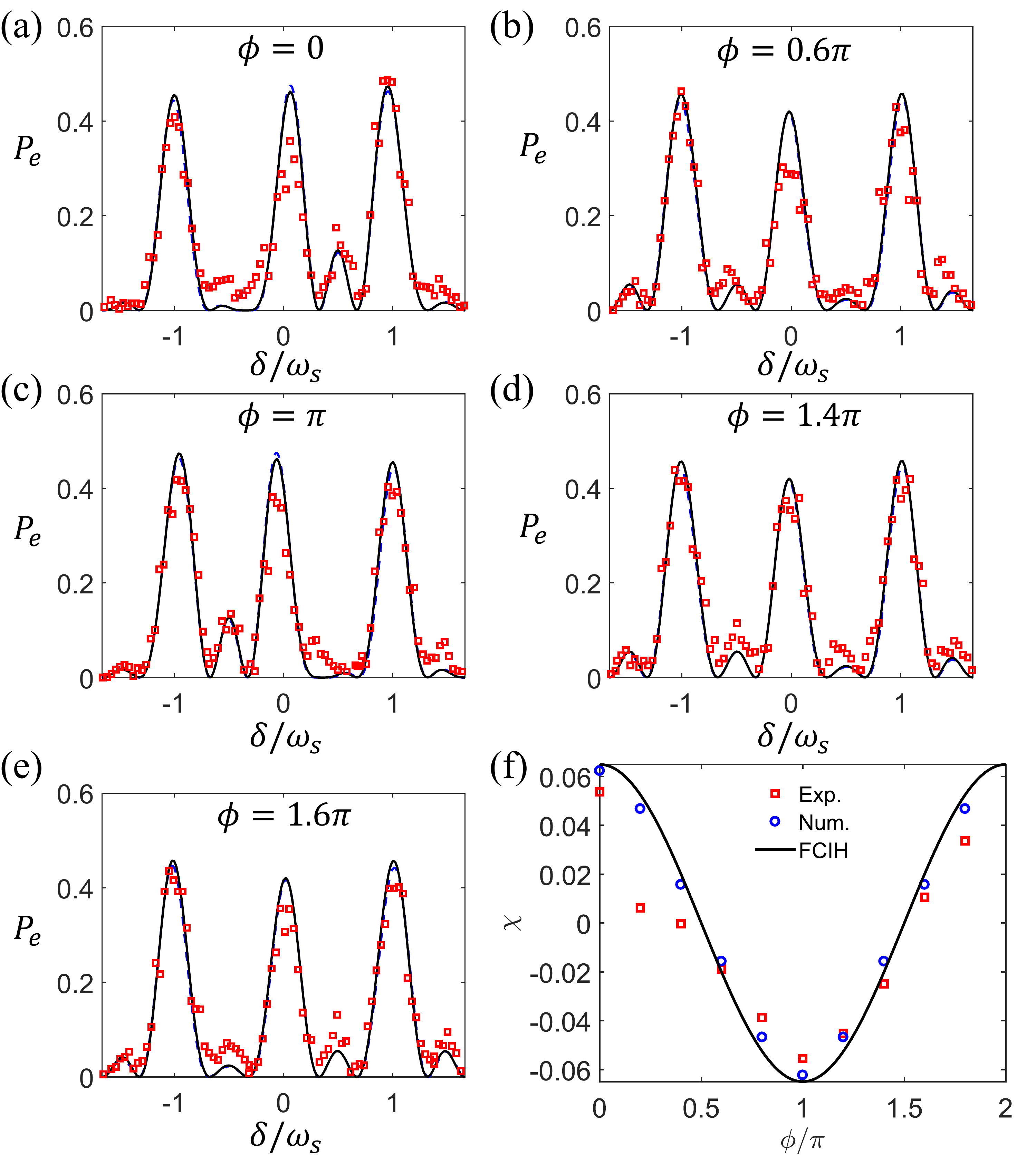}
	\caption{(a-e) The experimental (red dots), numerical (black solid lines), and FCIH (blue dashed lines) results of the Rabi spectrum for different initial phases at $g/2\pi=20.32$Hz. (f) The asymmetry contrast of the NIFBs for different initial phases.
 }\label{fig2}
\end{figure}

\textbf{\textit{Rabi Spectroscopy.--}} Based on the FCIH, the exited population $P_e(t)$ is determined by the interference of all the Floquet channels with effective detuning $\delta_k=\delta-k\omega_s$, so it can be explicitly expressed as 
\begin{eqnarray}\label{eq2}
P_e(t)=\bigg|\sum_k\Psi_k(t)\bigg|^2,\hspace{10px}\Psi_k(t)=-\frac{ig_k}{\Omega_k}  \sin \big[\frac{\Omega_k t}{2} \big] e^{\frac{-ik\omega_s t}{2}},
\end{eqnarray}
in which $\Omega_k=\sqrt{|g_k|^2+\delta_k^2}$ (see Supplemental Materials \cite{sup} for details). Definitely, the temperature effect can also be included by introducing the Boltzmann prefactor. Here, we want to emphasize that Eq.(\ref{eq2}) can not be derived by the perturbation theory which we have already tried. 

Under the stroboscopic measurement $t=nT_s$, the phase of $\Psi_k$ only depends on $g_k$. Considering $g_k\rightarrow g_k^{*}$ after transformation $\phi\rightarrow -\phi$, $\Psi_k$ will change into $-\Psi_k^*$, so the excited population $P_e(t)$ wll not be changed based on FCIH. Thus, the Rabi spectrum should be the same under the transformation $\phi\leftrightarrow-\phi$. On the other hand, if we implement the transformation $\phi\rightarrow\pi-\phi$, $g_k$ will change into $g_{-k}$, because the Bessel function fulfills $J_{-k}=(-1)^{k}J_k$. Then, the excitation population $P_e(t,\delta)$ will be transformed to $P_e(t,-\delta)$. Therefore, we can find another symmetry $\{\delta\leftrightarrow-\delta, \phi\leftrightarrow\pi-\phi\}$. The analysis of symmetries above is also suitable for the Ramsey spectrum, as demonstrated in Fig.\ref{fig1} (b-c).

The main contribution to the NIFB comes from the nearest neighbor Floquet channels. To simplify the problem, we focus on the NIFB between $k=0$ and $k=+1(-1)$ Floquet sidebands and name it the positive (negative) NIFB. Meanwhile, the driving amplitude in the experiment is adjusted to $A=1.47$ satisfying $J_0[A]\approx J_1[A]$, so that the peak of positive (negative) NIFB sits at $\delta=\omega_s/2$ and the phase difference between zero and first Floquet channels can be easily obtained $\Delta\phi=\frac{\omega_st}{2}+\phi$. Neglecting the influences of other Floquet channels, the excited population $P_e$ of positive and negative NIFBs under the stroboscopic measurement should be approximately proportional to $\cos^2{[\frac{\phi}{2}]}$ and $\sin^2{[\frac{\phi}{2}]}$, respectively. Then, their difference follows $\cos[\phi]$ while tuning the initial phase $\phi$.

The experimental results of the Rabi spectrum with different initial phases are presented in Fig.\ref{fig2}. Although the numerical results do not perfectly match the experimental data, many interesting features discussed above can still be recognized. At $\phi=0$ in Fig.\ref{fig2} (a), the high positive peak indicates strong constructive interference and the disappearance of the negative one hints the destructive interference. Such a strong interference effect can also be clearly observed at $\phi=\pi$ in Fig.\ref{fig2} (c) with the positive and negative NIFBs exchanged, like the Ramsey case in Fig.\ref{fig1} (b-c). When $\phi$ is tuned to other values, the NIFBs are messed up by the system's noise so can be hardly distinguishable as shown in Fig.\ref{fig2} (b,d-e). However, all the Rabi spectrum can still reflect the validity of both symmetries $\phi\leftrightarrow-\phi$ and $\{\delta\leftrightarrow-\delta, \phi\leftrightarrow\pi-\phi\}$. Furthermore, the surprisingly good coincidence between the analytic and numerical results supports the effectiveness of the FCIH \cite{sup}.

The positive and negative NIFBs are obscured by the system noise, but their difference can partially mitigate the influence of the noise. Thus, we introduce the asymmetry contrast
\begin{eqnarray}\label{eq3}
\chi=\frac{3}{\omega_s}\int^{2\omega_s/3}_{\omega_s/3} \left[P_e(\delta)-P_e(-\delta))\right]d\delta,
\end{eqnarray}
in which the integration from one trough to another can further weaken the noise's influence. As demonstrated in Fig.\ref{fig2} (f), the deviation between experimental and numerical results becomes much smaller except for small $\phi$. Even more strikingly, both of them match well with FCIH's prediction and follow the cosine function.

\begin{figure}[t]
	\centering
	\includegraphics[width=0.99\linewidth]{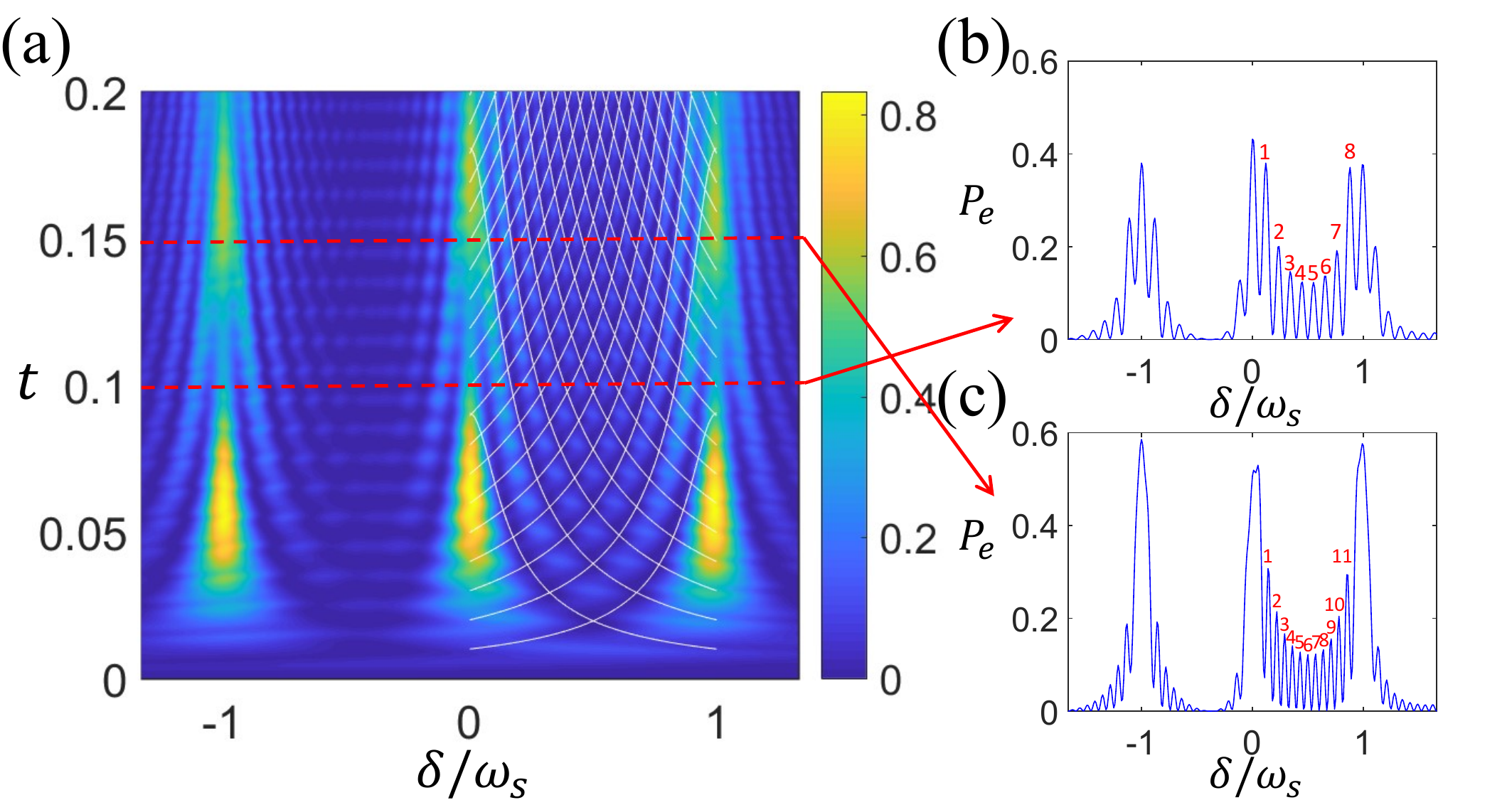}
	\caption{(a) The numerical results of the Rabi spectrum at $g/2\pi=20.32$Hz and $\phi=0$. The white lines denote the integer periods of $\Omega_0$ and $\Omega_1$. The red dashed lines highlight the spectrum at $t=0.1$ and $t=0.15$ which are shown in (b) and (c), respectively. The red numbers label the NIFBs.
 }\label{fig3}
\end{figure}

As the evolution time increases, due to the significant impact of decoherence and noise on our experimental platform, we are constrained to investigate the NIFBs through numerical and analytical methods. As shown in Fig.\ref{fig3} at $\phi=0$, the NIFBs emerge one by one as time goes on, resembling the interference pattern of two traveling waves. Based on the FCIH, the troughs can be approximately determined by the destructive interference of zeroth and first Floquet sidebands. In Fig.\ref{fig3}(a), the white lines show the integer periods of two Floquet channels: $t=\frac{n}{\Omega_0(\delta)}$
and $t=\frac{m}{\Omega_1(\delta)}$ ($n$,$m$ $\in \mathbb{Z}$). Then, the crossing points of white lines correspond to the commensurate value $\frac{\Omega_1}{\Omega_0}=\frac{m}{n}$ and indicate the local minimums. From Fig.\ref{fig3} (b-c), we can find that these minima do not equate to zero, a consequence of the additional phase factor $e^{\frac{-ik\omega_s t}{2}}$. Nevertheless, the NIFB can still stably stay between them, with its number increasing by two after each period $\frac1{\Omega_0(\omega_s/2)}$. However, we can find the number of NIFBs in Fig.\ref{fig3} (c) violates the prediction above, that is because of the merging of some NIFBs into the integer ones, as illustrated in Fig.\ref{fig3} (a). In Fig.\ref{fig3} (b), the height of the No. 1 and 8 NIFBs is comparable to that of the integer Floquet sidebands. This prompts the interesting question of whether the heights of NIFBs can surpass the integer ones in the experiment.

\begin{figure}[t]
	\centering
	\includegraphics[width=0.99\linewidth]{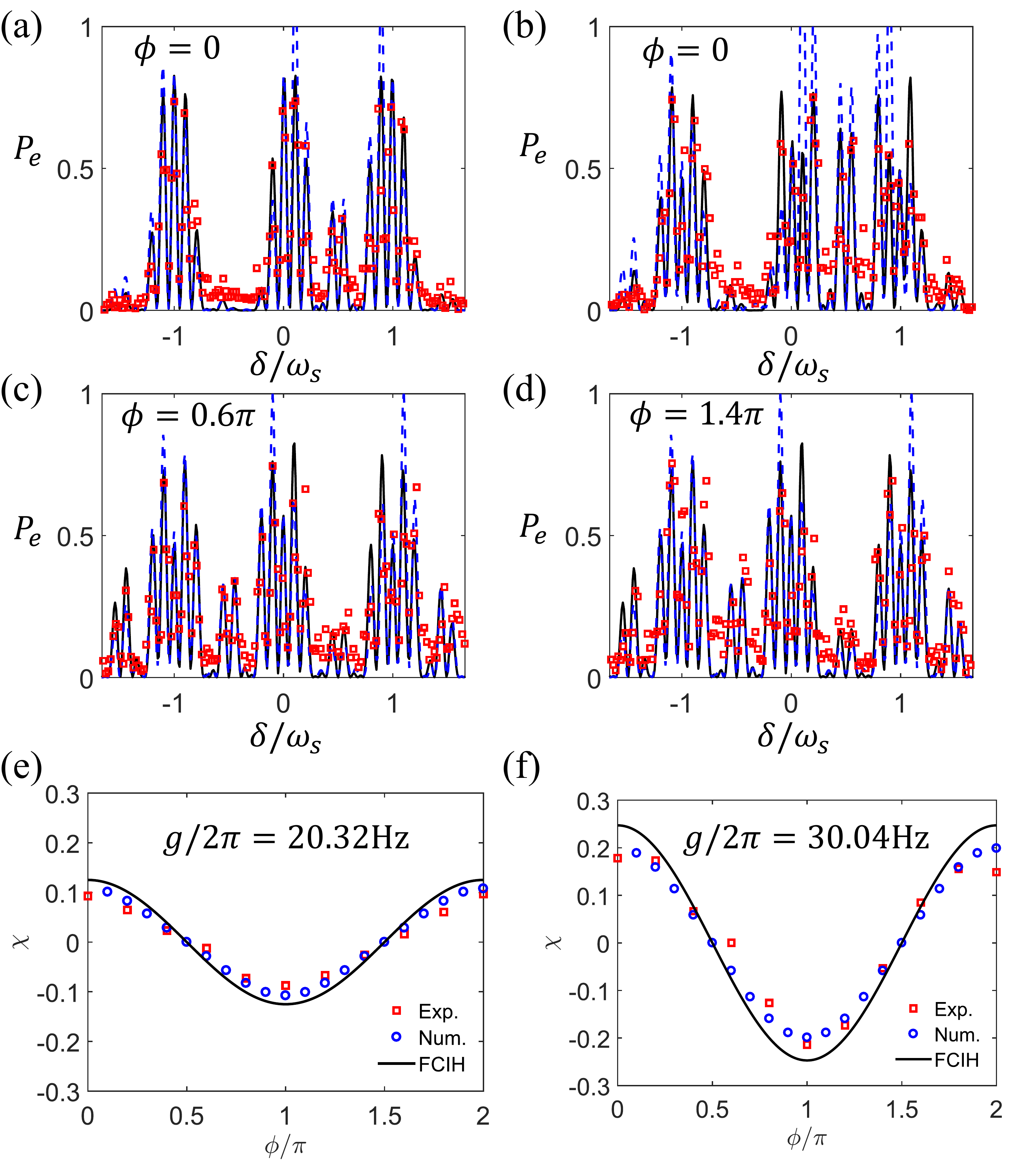}
	\caption{The experimental (red dots), numerical (black solid lines), and FCIH (blue dashed lines) results of the Ramsey spectrum for different initial phases at (a) $g/2\pi=20.32$Hz and (b-d) $g/2\pi=30.04$Hz. (e-f) The asymmetry contrast of the NIFBs for different initial phases.
 }\label{fig4}
\end{figure}

\textbf{\textit{ Ramsey spectroscopy--} }To enhance the quality of the spectrum, it is straightforward to consider Ramsey's method, which involves the insertion of a freely evolving period between two probing pulses \cite{ramsey}. The excited population $P_e$ can also be analytically obtained using the FCIH \cite{sup}. Differently, the $k$th Floquet wavefunction is expressed as
\begin{eqnarray}
\nonumber
    \Psi_k(t)&=&-\frac{2ig_{k}}{\Omega_k} e^{-i\frac{k\omega_s(2t_p+t_d)}{2})}\sin\frac{\Omega_k t_p}{2}\bigg(\cos{\frac{\Omega_k t_p}{2}} \cos{\frac{\delta_k t_d}{2}}\\
    &&-\frac{\delta_k}{\Omega_k}\sin{\frac{\Omega_k t_p}{2}} \sin{\frac{\delta_k t_d}{2}} \bigg),
\label{ramsey}
\end{eqnarray}
in which the phase only depends on $g_k$ while employing the stroboscopic measurement. Therefore, same as the Rabi spectroscopy, the symmetries $\phi\rightarrow -\phi$ and $\{\delta\leftrightarrow-\delta, \phi\leftrightarrow\pi-\phi\}$ still hold. 

The key idea of Ramsey spectroscopy is utilizing the interference to split the wide peak into several narrow ones, with linewidths proportional to $1/t_d$. In both Fig.\ref{fig1} (b-c) and Fig.\ref{fig4} (a-d), it becomes apparent that all the integer Floquet sidebands split into numerous narrow peaks. Notably, the Ramsey NIFBs with multiple peaks also emerge. These peaks exhibit high sensitivity to the initial phases and satisfy both symmetries, too.

As illustrated in Fig.\ref{fig1}(b) and Fig.\ref{fig4}(a), the numerical and FCIH results match well with the experimental data, particularly concerning the NIFBs. In comparison to the Rabi spectroscopy, the height of the Ramsey NIFBs is much larger ($P_e$=0.32). After that, we increase the probe laser power and adjust the Rabi frequency to $g/2\pi$=30.04Hz. As shown in Fig.\ref{fig4}(b-d), the intensity of the NIFBs is notably amplified and the symmetries still hold. In the experiment, the maximum peak height of the NIFB can reach 0.6 (see Fig.\ref{fig4}(b)), and the numerical simulation suggests that it could potentially exceed 0.9 with a continuous increase in $g$. However, immediate verification is impeded by the constraints of our experimental platform.

Indeed, despite the frequency interval of only 2Hz in our experiment, ensuring the distinctiveness of each Ramsey peak remains a challenge.  At large $g$, the numerical and experimental results still exhibit good agreement, whereas the FCIH results can only offer qualitative insights due to normalization issues. However, through integration, the asymmetry contrast $\chi$ significantly enhances the accuracy. As demonstrated in Fig.\ref{fig4} (e-f), both numerical and experimental results achieve a very good agreement with the FICH's prediction -- following the cosine function.

\textbf{\textit{Conclusion and discussion--}} The non-integer Floquet sideband spectroscopy is systematically investigated using both experimental and numerical methods. The proposed FCIH offers a quantitative description of the interference between different Floquet channels at small $g/\omega_s$. Building upon FCIH, we have identified two distinct symmetries associated with the initial phase, which have been validated through Rabi and Ramsey spectroscopy.  As the value of $g/\omega_s$ increases, the peaks of NIFBs grow significantly, even reaching a magnitude comparable to that of the integer Floquet sidebands.  Additionally, we have introduced a new parameter called asymmetry contrast $\chi$, which serves to mitigate the impact of noise and reflects the dependence on the initial phase. Furthermore, we have discussed the influence of spontaneous emission of the excited state and noise in the supplementary material \cite{sup}.

Our theoretical analysis presented here is not limited to the optical lattice clock, it can also be immediately applied to other platforms, such as NV-center, NMR, or trapped ion. Although the NIFBs discussed in our work pertain to single-body systems, we believe that their implications may also extend to quantum many-body systems. The discovery of NIFBs not only advances our understanding of Floquet physics but also holds the potential for enhancing quantum technologies based on Floquet engineering. 

\bibliography{ref}
\newpage
\newpage
\section{Supplementary Material}
\section{I. FCIH}
\subsection{A. Representation in the extended Hilbert space}
As mentioned in the main text, the Hamiltonian we considered is 
\begin{equation}
\label{Hphase}
\hat{H}=\frac{\hbar}{2}\bigg[(\delta+A\omega_s\cos\left(\omega_s t+\phi\right)) \hat{\sigma}_z+g\hat{\sigma}_x\bigg].
\end{equation}
Then, after implementing the unitary transformation $\hat{U}^{\dagger}(\hat{H}-i\hbar\frac{\partial}{\partial t})\hat{U}$ in which 
\begin{equation}
\hat{U}=\exp \left[-i \frac{A}{2}\sin{(\omega_s t+\phi)} \hat{\sigma}_z\right],
\end{equation}
we obtain the rotating Hamiltonian expressed as follows
\begin{equation}\label{ex}
\hat{H}_{r}=\frac{\hbar}{2}\bigg[\delta\hat{\sigma}_z+g\bigg(\sum_{n=-\infty}^{\infty} J_n[A] e^{in(\omega_st+\phi)} \hat{\sigma}^+ +h.c.\bigg)\bigg],
\end{equation}
where the Jacobi function $J_n[A]$ results from the Jacobi-Anger relations $e^{i z \sin \theta}=\sum_{n} J_{n}(z) e^{i n \theta}$.  According to the Floquet theory \cite{Eckardt2015}, the Hamiltonian Eq. (\ref{ex}) could be interpreted in an extended Hilbert space $\{\uparrow,\downarrow\} \bigotimes \{\boldsymbol{e^{\frac{i n\omega_s t}{2}\hat{\sigma}_z}}\}$, formed by the product of the Hilbert space of two energy level quantum system and the space of $2T_s$-periodically Fourier components. The wavefunction can be represented as 
\begin{equation}
|\psi(t)\rangle=\sum_{\alpha}c_{\alpha}|\alpha(t)
\rangle e^{-i\mathcal{E}_{\alpha}t/\hbar},
\end{equation}
so that 
\begin{equation}
i\hbar d_t|\alpha(t)
\rangle=(\hat{H}_{r} -\mathcal{E}_{\alpha})|\alpha(t)
\rangle.
\end{equation}
Based on the Floquet theory, we have the condition $|\alpha(t) \rangle=|\alpha(t+T_s)\rangle$, so we can define the basis vector of Floquet modes as 
\begin{equation}
|\alpha m\rangle\rangle=\frac{1}{2T_s}\int^{2T_s}_0  e^{\frac{im\omega_s t}2} |\alpha(t)\rangle dt.
\end{equation}
Then the time evolution of the state $|\psi(t)\rangle$ can be expressed as:
\begin{eqnarray}\label{psitime}
|\psi(t)\rangle&=
\sum\limits_{\alpha m} c_{\alpha} e^{-\frac{i\mathcal{E}_{\alpha m}t}{\hbar}} |\alpha m\rangle \rangle,
\end{eqnarray}
and the Schr\"{o}dinger equation can be written as 
\begin{equation}
\hat{Q}|\alpha m \rangle \rangle=\mathcal{E}_{\alpha m}|\alpha m \rangle \rangle
\end{equation}
where the quasi-energy $\mathcal{E}_{\alpha m}=\mathcal{E}_{\alpha}+m\hbar \omega_s$ playing the role as eigen-energy and operator $\hat{Q}=\hat{H}_r-i\hbar d_t$ as the static Hamiltonian. 

In extended Hilbert space, the operator $\hat{Q}$ can be expressed as following matrix form:
\begin{equation}
\begin{bmatrix}\label{mat}
\ddots&\vdots&\vdots&\vdots&\vdots&\vdots& \\
\cdots&\hat{H}_{2;2}&\hat{H}_{1;2}&\hat{H}_{0;2}&\hat{H}_{-1;2}&\hat{H}_{-2;2}&\cdots\\
\cdots&\hat{H}_{2;1}&\hat{H}_{1;1}&\hat{H}_{0;1}&\hat{H}_{-1;1}&\hat{H}_{-2;1}&\cdots\\
\cdots&\hat{H}_{2;0}&\hat{H}_{1;0}&\hat{H}_{0;0}&\hat{H}_{-1;0}&\hat{H}_{-2;0}&\cdots\\
\cdots&\hat{H}_{2;-1}&\hat{H}_{1;-1}&\hat{H}_{0;-1}&\hat{H}_{-1;-1}&\hat{H}_{-2;-1}&\cdots\\
\cdots&\hat{H}_{2;-2}&\hat{H}_{1;-2}&\hat{H}_{0;-2}&\hat{H}_{-1;-2}&\hat{H}_{-2;-2}&\cdots\\
&\vdots&\vdots&\vdots&\vdots&\vdots&\ddots\\
\end{bmatrix}
\end{equation}
with the 2x2 matrix $\hat{H}_{m;n}$ defined as 
\begin{equation}
\begin{aligned}
&\hat{H}_{m;n}=\langle\langle \alpha' m| \bar{Q} | \alpha n \rangle \rangle\\
&=\frac{1}{2T_s}\int_{0}^{2T_s} dt \; e^{\frac{-i m\omega_s t}{2} \hat{\sigma}_z} ( \hat{H}_{r}(t)-i\hbar d_t) e^{\frac{i n\omega_s t}{2}\hat{\sigma}_z}\\
&=\hbar
\begin{cases}
\begin{bmatrix}
\frac{\delta-k \omega_s}{2}&\frac{g}{2}J_{-k} e^{-ik\phi} \\
\frac{g}{2}J_{-k} e^{ik\phi}&-\frac{\delta-k \omega_s}{2}
\end{bmatrix}
&(m=n=-k)\\
\begin{bmatrix}
0&\frac{g}{2} J_{-k}e^{-ik\phi}\\
\frac{g}{2} J_{-k}e^{ik\phi}&0
\end{bmatrix}
&(m+n=-2k,m\neq n)\\
0&(m+n=-2k+1)

\end{cases}, k\in \mathbb{Z}.
\end{aligned}
\end{equation}
The dimension of the matrix (\ref{mat}) is infinite, making it still hard to handle. However, if we observe the block-diagonal terms:
\begin{equation}\label{effhami}
\hat{H}_{k}=
\begin{bmatrix}
\frac{\delta-k \omega_s}{2}&\frac{g}{2}J_{-k} e^{-ik\phi} \\
\frac{g}{2}J_{-k} e^{ik\phi}&-\frac{\delta-k \omega_s}{2}
\end{bmatrix},
\end{equation}
, it is exactly the effective Hamiltonian of kth Floquet channel under RSBA \cite{MoJuanYin2020} with eigenvalues and eigenvectors expressed as:
\begin{equation}
\begin{aligned}
\mathcal{E}_{k}=&
\begin{bmatrix}
\frac{\Omega_k}{2}&0\\
0&-\frac{\Omega_k}{2}
\end{bmatrix}\\
R_k=&
\begin{bmatrix}
\cos{\gamma_k}&\sin{\gamma_k}\\
-\sin{\gamma_k}&\cos{\gamma_k}
\end{bmatrix}, \gamma_k=\mathrm{ArcTan}(\delta_k/g_k).
\end{aligned}
\end{equation}

\subsection{B. FCIH on Rabi and Ramsey Spectroscopy}
The atoms are prepared in the ground state $|\downarrow\rangle$, so the evolution of the state in the $k$th Floquet channel is
\begin{equation}
\label{transmat}
|\psi_k{(t)} \rangle=
\begin{bmatrix}
-\frac{ig_k}{\Omega_k}  \sin[\frac{\Omega_k}{2}t]\\
\cos(\frac{\Omega_k}{2}t)+\frac{i\delta_k}{\Omega_k}\sin(\frac{\Omega_k}{2}t) 
\end{bmatrix}.
\end{equation}
Turning back to the original Hilbert space, the Fourier factors should be included in the wavefunction as:
\begin{equation}
\label{transmat}
|\psi_k{(t)} \rangle=
\begin{bmatrix}
-\frac{ig_k}{\Omega_k} \sin[\frac{\Omega_k}{2}t] e^{-\frac{ik\omega_s t}{2}}\\
\big[\cos(\frac{\Omega_k}{2}t)+\frac{i\delta_k}{\Omega_k}\sin(\frac{\Omega_k}{2}t) \big]e^{\frac{ik\omega_s t}{2}}.
\end{bmatrix}
\end{equation}
As we proposed in the main text, the basic idea of FCIH is the superposition of wavefunction in different Floquet channels, so the expression of the excited probability of Rabi spectroscopy is written as follows:
\begin{equation}\label{modrfsba}
\begin{aligned}
P_e(t)
&=\bigg|\sum_k\bigg(-\frac{ig_k}{\Omega_k}  \sin \big[\frac{\Omega_k t}{2} \big]e^{\frac{-ik\omega_s t}{2}} \bigg)\bigg|^2
\end{aligned}.
\end{equation}

The derivation of Ramsey spectroscopy is also straightforward. The Ramsey process can be viewed as three consecutive Rabi processes, with the second Rabi process having $g=0$ \cite{ramsey}. The transfer matrix of the Rabi process in $k$th Floquet channel is
\begin{equation}
\begin{aligned}
&M(t,g,k)=\\
&\begin{bmatrix}
\cos(\frac{\Omega_k}{2}t)-\frac{i\delta_k}{\Omega_k}\sin(\frac{\Omega_k}{2}t) 
&-\frac{ig_{k}}{\Omega_k}  \sin[\frac{\Omega_k}{2}t] \\
-\frac{ig_{k}^{*}}{\Omega_k}  \sin[\frac{\Omega_k}{2}t] &
\cos(\frac{\Omega_k}{2}t)+\frac{i\delta_k}{\Omega_k}\sin(\frac{\Omega_k}{2}t) 
\end{bmatrix}\\
\end{aligned}
\end{equation}
so after the Ramsey process with transfer matrix $M(t_p,g,k)M(t_d,0,k)M(t_p,g,k)$, the wavefunction changes from ground state $|\downarrow\rangle$ to
\begin{footnotesize}
	\begin{equation}
	|\psi_k{(t)} \rangle=\begin{bmatrix} 
	-\frac{2ig_{k}}{\Omega_k} \sin\frac{\Omega_k t_p}{2} \bigg(\cos{\frac{\Omega_k t_p}{2}} \cos{\frac{\delta_k t_d}{2}}-\frac{\delta_k}{\Omega_k}\sin{\frac{\Omega_k t_p}{2}} \sin{\frac{\delta_k t_d}{2}} \bigg)\\
	-\frac{|g_{k}|^2}{\Omega_k^2} e^{-\frac{i \delta_k t_d}{2}}\sin^2 {\frac{\Omega_k t_p}{2}}+e^{\frac{i\delta_k t_d}{2}} \bigg(\cos{\frac{\Omega_k t_p}{2}}+\frac{i \delta_k}{\Omega_k} \sin{\frac{\Omega_k t_p}{2}} \bigg)^2
	\end{bmatrix}.
	\end{equation}
\end{footnotesize}
Thus, with the help of FCIH, the excited probability of Ramsey spectroscopy is
\begin{equation}
\begin{split}
P_e=&\bigg|\sum_{k} -\frac{2ig_{k}}{\Omega_k} e^{-i\frac{k\omega_s(2t_p+t_d)}{2})}\sin\frac{\Omega_k t_p}{2}\\
&\bigg(\cos{\frac{\Omega_k t_p}{2}} \cos{\frac{\delta_k t_d}{2}}-\frac{\delta_k}{\Omega_k}\sin{\frac{\Omega_k t_p}{2}} \sin{\frac{\delta_k t_d}{2}} \bigg) \bigg|^2,
\end{split}
\label{ramseyexpression}
\end{equation}
where the phase $e^{-ik (\phi+\frac{\omega_s(2t_p+t_d)}{2})}$ governing the interference between different Floquet channels.

\begin{figure}[t]
	\includegraphics[width=0.98\linewidth]{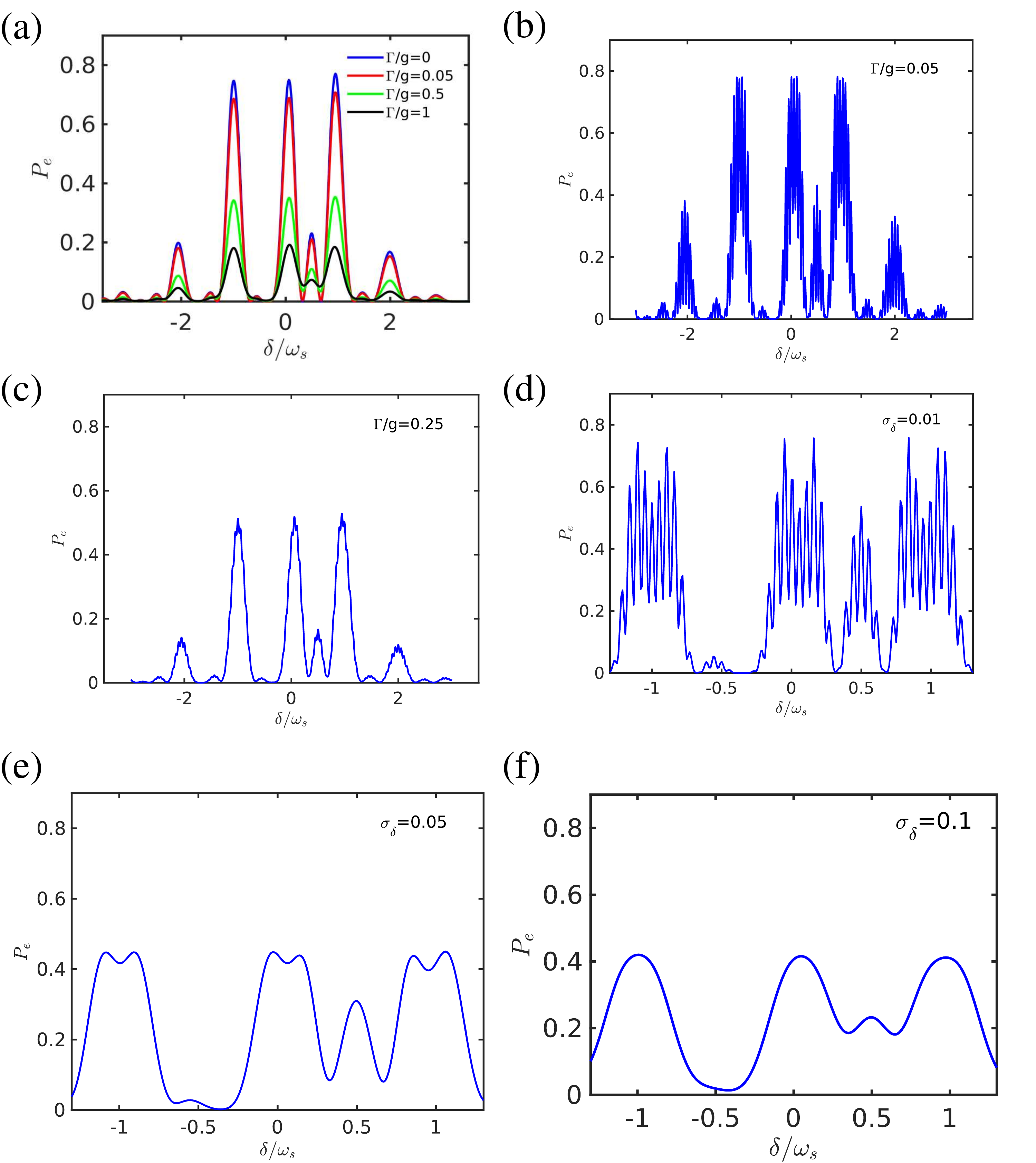}
	\caption{ (a) Rabi spectra at different $\Gamma$. (b-c) Ramsey spectra at different $\Gamma$. (d-f): Ramsey Spectra under noise with different standard deviation $\sigma_{\delta}$. The other system parameters are $A=1.47$, $g/\omega_s=0.2$, $t_p=3T_s$, and $t_d=15T_s$}
	\label{noisepeak} 
\end{figure}
\section{II. Robustness of NIFBs}
Although the NIFBs have been clearly observed in the experiment, it is still worthwhile to theoretically discuss the impact of disturbing factors, such as spontaneous emission and noise. Firstly, we consider the influence of spontaneous emission. We employ the master equation in Lindblad form \cite{mastere}:
\begin{equation}\label{mastereq}
\frac{d}{dt}\hat{\rho}_A(t)=\frac{1}{i \hbar}[\hat{H},\hat{\rho_A}(t)]+\sum_i\bigg(\hat{L}_i \hat{\rho}_A(t) \hat{L}_i^{\dagger}-\frac{1}{2}[\hat{L}_i^{\dagger} \hat{L}_i,\hat{\rho}_A(t)] \bigg)
\end{equation}
where $\hat{\rho}_A$ is the density matrix of the system, and $\hat{L}_i$ is the Lindblad term representing the influence of the environment on the system. The effect of spontaneous emission we considered involves the transition from the excited state to the ground state, so the Lindblad term is expressed as:
\begin{equation}
\hat{L}=\sqrt{\Gamma} |g \rangle \langle e|
\end{equation}
in which $\Gamma$ is the spontaneous emission rate. The numerical results are shown in Fig.\ref{noisepeak}(a-c). We can find that spontaneous emission doesn't change the position of peaks, but rather causes a gradual decrease in the height and fine structure of each peak until only the envelopes remain. However, the significant asymmetric feature is pretty robust.

For the OLC platform, the frequency noise can affect the longitudinal term $\delta \hat{\sigma}_z$. Here, we utilize Gaussian distributed noise to evaluate its impact on Ramsey Spectra, and the averaged excited probability should be:
\begin{equation}
\overline{P_e(\delta,g,\sigma_{\delta})}
=\frac{1}{\sqrt{2\pi\sigma_{\delta}}} \int P_e(y,g) \exp\bigg[-\frac{(y-\delta)^2}{2\sigma_{\delta}^2} \bigg] dy.
\label{noise}
\end{equation}
The corresponding numerical results of the Ramsey Spectra are shown in Fig.\ref{noisepeak}(d-f). We can see that the frequency noise smooths the sharp peaks and weakens their strength. However, the asymmetry of the Ramsey spectra is still distinguishable. 

Although we can not consider all the influences, the analysis of spontaneous emissions and frequency noise already reflects the robustness of NIFBs.

\end{document}